\pdfoutput=1
\documentclass[fleqn,usenatbib]{mnras}
\usepackage{amsmath}	

\usepackage{mathptmx}
\usepackage{txfonts}

\usepackage[T1]{fontenc}

\usepackage{graphicx}	
\usepackage{amssymb}	

\title[SDSS~J1152+0248: An eclipsing double WD]{SDSS~J1152+0248: An eclipsing double white dwarf from the \textit{Kepler} \textit{K2} campaign}

\author[N. Hallakoun et al.]{
N. Hallakoun,$^{1,2}$\thanks{E-mail: \href{mailto:naama@wise.tau.ac.il}{naama@wise.tau.ac.il}}
D. Maoz,$^{1}$
M. Kilic,$^{3}$
T. Mazeh,$^{1}$
A. Gianninas,$^{3}$
E. Agol,$^{4}$
K. J. Bell,$^{5}$
\newauthor{
S. Bloemen,$^{6}$
W. R. Brown,$^{7}$
J. Debes,$^{8}$
S. Faigler,$^{1}$
I. Kull,$^{1}$
T. Kupfer,$^{6}$
A. Loeb,$^{9}$}
\newauthor{
B. M. Morris,$^{4}$
and F. Mullally$^{10}$}
\\
\\
$^{1}$School of Physics and Astronomy, Tel-Aviv University, Tel-Aviv 6997801, Israel\\
$^{2}$European Southern Observatory, Karl-Schwarzschild-Stra{\ss}e 2, 85748 Garching, Germany\\
$^{3}$Department of Physics and Astronomy, University of Oklahoma, 440 W. Brooks St, Norman, OK 73019, USA\\
$^{4}$Department of Astronomy, Box 351580, University of Washington, Seattle, WA 98195, USA\\
$^{5}$Department of Astronomy, University of Texas at Austin, Austin, TX 78712, USA\\
$^{6}$Department of Astrophysics, IMAPP, Radboud University Nijmegen, PO Box 9010, NL-6500 GL Nijmegen, the Netherlands\\
$^{7}$Smithsonian Astrophysical Observatory, 60 Garden St, Cambridge, MA 02138, USA\\
$^{8}$Space Telescope Science institute, 3700 San Martin Dr., Baltimore, MD 21218, USA\\
$^{9}$Department of Astronomy, Harvard University, 60 Garden St, Cambridge, MA 02138, USA\\
$^{10}$SETI Institute/NASA Ames Research Center, Moffet Field, CA 94035, USA
}

\date{Accepted XXX. Received YYY; in original form ZZZ}

\pubyear{2016}

\begin{document}
\label{firstpage}
\pagerange{\pageref{firstpage}--\pageref{lastpage}}
\maketitle

\begin{abstract}
We report the discovery of the sixth known eclipsing double white dwarf (WD) system, SDSS~J1152+0248, with a $2.3968\pm0.0003$\,h orbital period, in data from the \textit{Kepler} Mission's \textit{K2} continuation. Analysing and modelling the \textit{K2} data together with ground-based fast photometry, spectroscopy, and radial-velocity measurements, we determine that the primary is a DA-type WD with mass $M_1=0.47\pm0.11\,M_{\sun}$, radius $R_1=0.0197\pm0.0035\,R_{\sun}$, and cooling age $t_1=52\pm36$\,Myr. No lines are detected, to within our sensitivity, from the secondary WD, but it is likely also of type DA. Its central surface brightness, as measured from the secondary eclipse, is $0.31$ of the primary's surface brightness. Its mass, radius, and cooling age, respectively, are $M_2=0.44\pm0.09 M_{\sun}$, $R_2=0.0223^{+0.0064}_{-0.0050}\,R_{\sun}$, and $t_2=230\pm100$\,Myr. SDSS~J1152+0248 is a near twin of the double-lined eclipsing WD system CSS~41177.
\end{abstract}



\section{Introduction}\label{sec:Intro}
White dwarfs (WDs) in general, and WDs in binary systems in particular, are important in a broad range of astrophysical contexts. To name some: for understanding the final stages of stellar and close-binary evolution \citep[e.g.][]{Benvenuto+1998,Holberg+2012}, as progenitors of hot subdwarfs \citep[e.g.][]{Han+2002,Han+2003} and Type-Ia supernovae \citep[e.g.][]{Maoz+2014}; as sources of gravitational waves \citep[e.g.][]{Nelemans_2009,Amaro-Seoane+2013};  and even as possible hosts of extrasolar planets on which life could potentially be detected \citep{Agol_2011,Loeb+2013}. Eclipsing WD systems can be particularly revealing in that they permit direct measurements of the radii and masses of their constituents.

Over the last years, the number of known binary systems with a WD component has increased significantly. There are more than 70 known double WDs \citep{Marsh_2011,Kaplan+2012,Brown+2013,Gianninas+2015}. Since the 2010 discovery of the first eclipsing double WD binary, NLTT~11748 \citep{Steinfadt+2010}, four more such systems have been discovered \citep{Parsons+2011,Vennes+2011,Brown+2011,Kilic+2014}. Among them, there is one known case of a double-lined eclipsing double WD system, CSS~41177, in which full model-independent derivation of the system parameters is possible \citep{Bours+2014}.

In this paper, we present the discovery of the sixth known eclipsing double WD system, SDSS~J115219.99+024814.4 (hereafter SDSS~J1152+0248). SDSS~J1152+0248 was identified as a $g{=}18.35$\,mag WD candidate in the Sloan Digital Sky Survey (SDSS) based on colour selection \citep{Girven+2011}. It was included as a target for the 2-Wheeled \textit{Kepler} continuation mission (\textit{K2}) as a part of a program targeting $\sim 150$ WDs \citep{Kilic+2013}. As detailed below, photometry of the target, obtained during `Campaign 1' of the \textit{K2} mission, revealed periodic ($P=2.4$\,h) primary and secondary eclipses. Follow-up spectroscopy confirmed the primary as a DA WD, and showed radial-velocity modulation with the same periodicity, while fast photometry revealed the details of the primary and secondary eclipses. Below, we describe these observations, their analysis, and our modelling of the system and its parameters.

\section{Observations}
\subsection{Photometry: \textit{K2}}
SDSS~J1152+0248 was observed in \textit{K2} Campaign 1, and was given the Ecliptic Plane Input Catalogue (EPIC) number 201649211. The campaign lasted from 2014 May 30 to 2014 August 21, and covered a field in the North Galactic Cap. \textit{Kepler} data are divided into `cadences'. Our target's data have a `long cadence' in which 270 frames are coadded on board, before the photometry of each target is downlinked to Earth, resulting in an integrated exposure time per epoch of $1766$\,s, or about half an hour.\footnote{See http://archive.stsci.edu/kepler/manuals/Data\_Characteristics.pdf.}

Analysis of the light curve, described in detail in Section~\ref{sec:K2}, below, revealed periodic ($2.4$\,h) primary and secondary eclipses, each of duration of 30\,min in the phase-folded light curve. The primary eclipse depth was $\sim 2\%$. The similarity between the eclipse duration and the observing cadence suggested much deeper and briefer true eclipses, indicating a WD companion to the candidate primary WD.

\subsection{Spectroscopy: WHT}\label{sec:WHT}
A spectrum of the primary WD was obtained on 2015 March 16 with the Intermediate-dispersion Spectrograph and Imaging System (ISIS) on the $4.2$-m William Herschel Telescope (WHT) in La Palma, Spain. A $1300$\,s exposure using the two channels of the instrument covered the ranges of $3600$ to $5200$\,\AA\ and $5600$ to $8000$\,\AA\ at $0.8$\,\AA\ and $0.9$\,\AA\ resolution, respectively. This spectrum confirmed SDSS~J1152+0248 as a DA-type WD (i.e. with a hydrogen atmosphere).

\subsection{Photometry: Wise}
To investigate the suspected short eclipse duration, we obtained unfiltered imaging photometry using the PI camera on the $1$-m telescope of the Wise Observatory in Israel, on the nights of the 2015 March 17 and 22, covering about two full cycles  of the system on each night. The observational cadences were $300$\,s and $30$\,s on the first and second night, respectively. These observations confirmed that the eclipse durations are of order $1$\,min, with primary and secondary eclipse depths of about 50\% and 10\%, respectively.

\subsection{Spectroscopy: MMT}\label{sec:MMT}
Additional spectra of the system were obtained using the Blue Channel Spectrograph on the Multiple Mirror Telescope (MMT) in Mt. Hopkins, Arizona, to search for radial-velocity (RV) variations in the primary WD, and for spectral signatures of the putative secondary WD. Thirteen exposures of $360$\,s each were taken with the 832 line mm$^{-1}$ grating and a 1\,arcsec slit on the night of 2015 March 24, covering the range $3600$ to $4500$\,\AA\ with $1.0\,\AA$ resolution. As detailed in section~\ref{sec:RV}, the spectra revealed RV variations at the expected period and phase, with semi-amplitude $K_1 \approx 110$\,km\,s$^{-1}$. These spectra did not reveal any sign of the secondary, neither in the high-order Balmer lines (as might be expected if the secondary were another, bright enough, DA WD), nor in the presence of any other lines in the spectrum, to an equivalent-width limit of $\sim 0.4\,\AA$. As a further test, we obtained additional spectra of the H$\alpha$ region of the spectrum, as the narrow non-LTE (NLTE) core present in the H$\alpha$ line of WDs permits resolving close spectral components. Four exposures, of $1080$\,s each, were taken with 1200 line mm$^{-1}$ grating and a 1\,arcsec slit, on the night of 2015 March 27, covering the range $5855$ to $7165$\,\AA\ with $1.5\,\AA$ resolution, and were obtained approximately at the times of quadratures of the cycle (as predicted from our \textit{Kepler} ephemeris for this system and from the RV curve obtained 3 nights before), in order to give the maximal spectral separation between the two components (if present). The spectrum reveals only a single H$\alpha$ component. However, based on the system parameters we derive below (see Section~\ref{sec:Combined}), we would not have expected to detect the Balmer lines from the secondary, given the signal-to-noise ratio of the spectrum. Although, in principle, the secondary could be a DC-type WD (a cool helium-atmosphere WD with a featureless optical continuum), this would be inconsistent with the lack of dilution we find in the primary's Balmer lines (see Section~\ref{sec:Atmospheric}). Furthermore, we will show that the secondary has a mass $\sim 0.45 M_{\sun}$, and \citet{Bergeron+2011} have shown that helium atmosphere (DB-type)  WDs rarely have masses $\lesssim 0.5 M_{\sun}$. Finally, we show in Section~\ref{sec:Combined} that the results of our radial velocity estimates are likely affected by the presence in the spectrum of lines from the secondary WD. We conclude that the secondary is likely a DA WD, one whose lines and kinematics could be measured with deeper observations, which would then make this system the second known double-lined eclipsing WD system.

\subsection{Photometry: APO}
To probe the structure of the eclipses for the purpose of constraining the system's geometry, fast photometry was performed on the night of 2015 April 15 with the Agile photometer on the ARC $3.5$-m telescope at Apache Point Observatory (APO) in New Mexico. A continuous sequence of $10$\,s exposures spanned two full cycles of the system. Observations used a BG40 filter, which is a broad-band filter covering $\sim 3400-6000\,\AA$. After dark subtraction and flat correction, we retrieved aperture photometry on SDSS~J1152+0248 and two nearby comparison stars. We chose aperture radii of 2\,arcseconds which minimized the variance of the out-of-eclipse light curve. We constructed the light curve by dividing the flux from the target star by a linear combination of the fluxes of the comparison stars, and decorrelated trends with airmass.

\subsection{Photometry: McDonald}
Additional fast photometry was obtained on the night of 2015 April 18 with the ProEM Camera on the $2.1$-m Otto Struve telescope at the McDonald Observatory in Texas. Again, $10$\,s exposures with the BG40 filter (similar to the BG40 filter used at APO, see above) covered two full system cycles. We used the IRAF script \texttt{ccd\_hsp} \citep{Kanaan+2002} to perform aperture photometry, and \texttt{WQED} \citep{Thompson+2013} to divide the target flux by the flux of a comparison star in order to correct for transparency variations, and to compute the Barycentric correction.

\section{Data analysis and modeling}
\subsection{\textit{K2} light curve and orbital period}\label{sec:K2}
\begin{figure}
\includegraphics[width=\columnwidth]{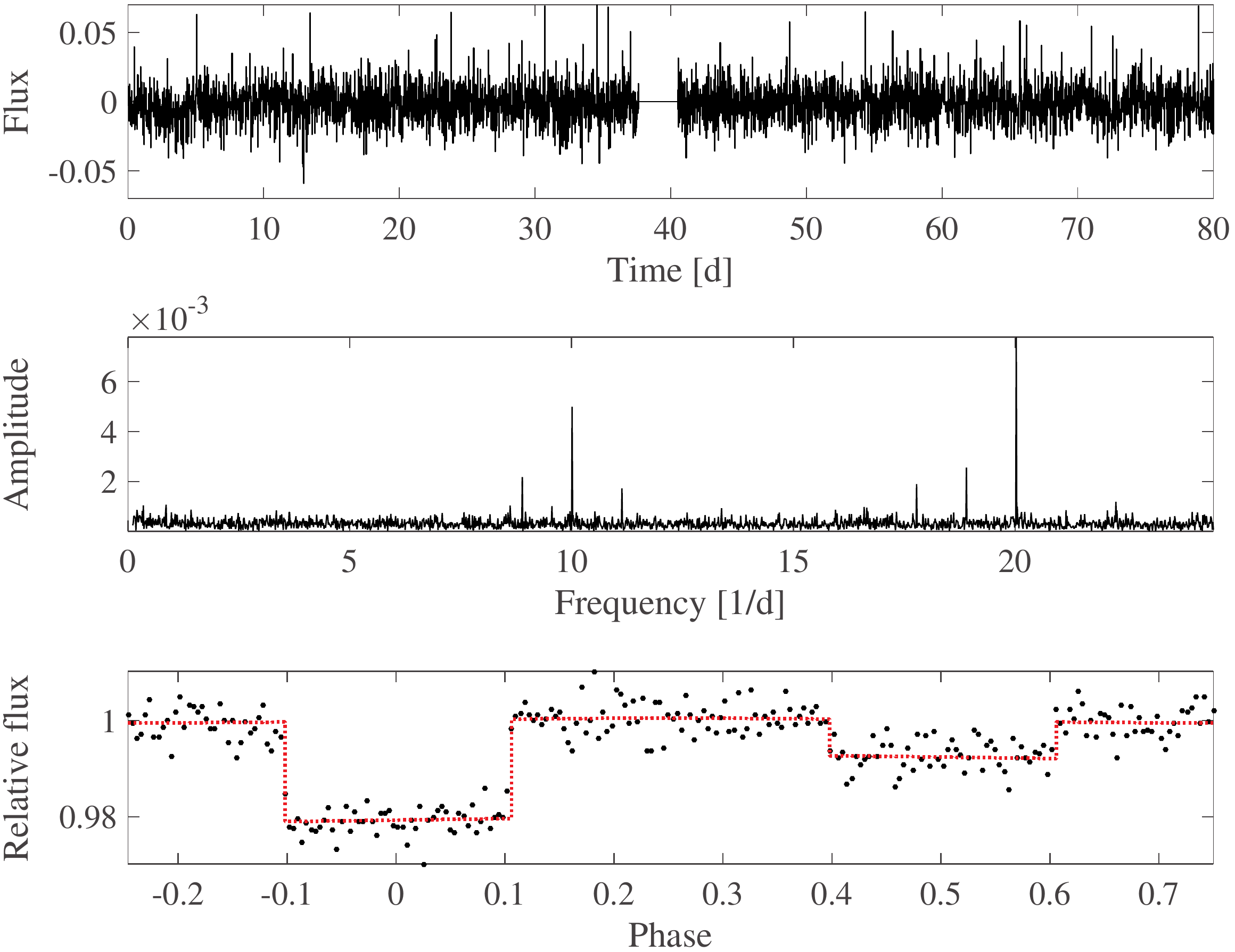}
\caption{\textit{K2} data for SDSS~J1152+0248. Top: reduced and detrended light curve; middle: FFT amplitude spectrum; bottom: phase-folded and binned light curve. Dashed red curve is best-fitting eclipse plus relativistic beaming model (see Section~\ref{sec:Model}).}
\label{fig:K2}
\end{figure}

The periodic eclipses in SDSS~J1152+0248 were discovered in the process of searching for periodicities in the \textit{K2} data for WDs. Without two of its original reaction wheels, the images of the \textit{K2} mission are less stable than those of the original \textit{Kepler} mission.

\citet{Vanderburg+2014} developed a self-flat-fielding photometry technique to reduce the effects of spacecraft roll on the \textit{K2} data. We used manually defined target apertures and the \texttt{kepsff} package under \texttt{PyKE} \citep{Still+2012} to reduce the \textit{K2} data. Self flat fielding reduced most of the spacecraft roll effects, but with remaining long-timescale variations. Additional artefacts in the data include small gaps of order an hour caused by the removal of low-quality cadences induced by maintenance procedures (such as angular momentum desaturation of the reaction wheels); a single $\sim 3$\,day-long break in the data in the middle of the campaign, for downlinking collected science data to Earth; and an unevenly sampled grid that results from the conversion of Universal Time (UTC) to barycentre-corrected Julian date (BJD) in the middle of each cadence.

To deal with these problems, we further processed each WD light curve through the following steps. The reduced light curve was reassigned to an evenly spaced grid using a nearest-neighbour interpolation, and gaps in the data were replaced by zeros. Post-reduction trends that remained in the light curve were removed by breaking up the data into $5$\,d intervals, and subtracting a linear fit from each interval, resulting in a trend-free light curve with zero mean. Data points deviating by more than 3 standard deviations from the mean were replaced by zeros. A Fourier amplitude spectrum of the light curve was then calculated and used to get an initial estimate of the period. The period was fine-tuned by inspecting the phase-folded light curve and adjusting the period until the least scatter was achieved. The scatter was determined by measuring the resulting modulation amplitude over a range of values close to the found period. The period errors were defined as the $\pm \sigma$ intervals of the mean of a Gaussian fit of the scatter. Finally, the period-folded light curve was binned into 250 bins, each containing the median of its sample points. The error in each bin was estimated as $1.48$ times the median absolute deviation around the median, divided by $\sqrt{n_{\textrm{bin}}}$, where $n_{\textrm{bin}}$ is the number of points in that bin.

Fig.~\ref{fig:K2} shows, for SDSS~J1152+0248: the detrended \textit{K2} light curve; the Fourier amplitude spectrum, with a clear peak at $10$\,d$^{-1}$, and a strong harmonic at 20\,d$^{-1}$; and the phase-folded light curve, revealing a primary and secondary eclipse, with a period of $2.3968\pm0.0003$\,h. The second harmonic arises partly from the secondary eclipse, which together with the primary eclipse introduces into the Fourier spectrum a double frequency, and partly from the highly non-sinusoidal form of the variations. The additional peaks in the Fourier spectrum result from higher harmonics, beyond the Nyquist frequency, that get reflected and shifted back into the spectrum. As already noted, the eclipse duration of 30\,min in the \textit{K2} data, similar to the \textit{K2} time resolution per individual measurement, indicates a much briefer and deeper true eclipse that has been diluted by the \textit{K2} cadence.

\subsection{Eclipse modelling from fast photometry data}\label{sec:Model}

We next combine the APO and McDonald fast photometry data to obtain time-resolved coverage of the primary and secondary eclipses. We then model these data to constrain some of the system's geometric parameters.

To obtain reliable estimates of photometric errors, each light curve, from APO and from McDonald, was assigned a constant error of $1.48$ times the median absolute deviation around the median of points outside of the eclipses. This resulted in errors of $5\%$ for the APO data and $3\%$ for McDonald. We tested for correlated errors among consecutive data points by calculating the auto-correlation function (ACF) for each of the light curves. Both ACFs fall to near zero in one time step, showing that any correlated errors are at a level well below the random errors.

\begin{figure}
\includegraphics[width=\columnwidth]{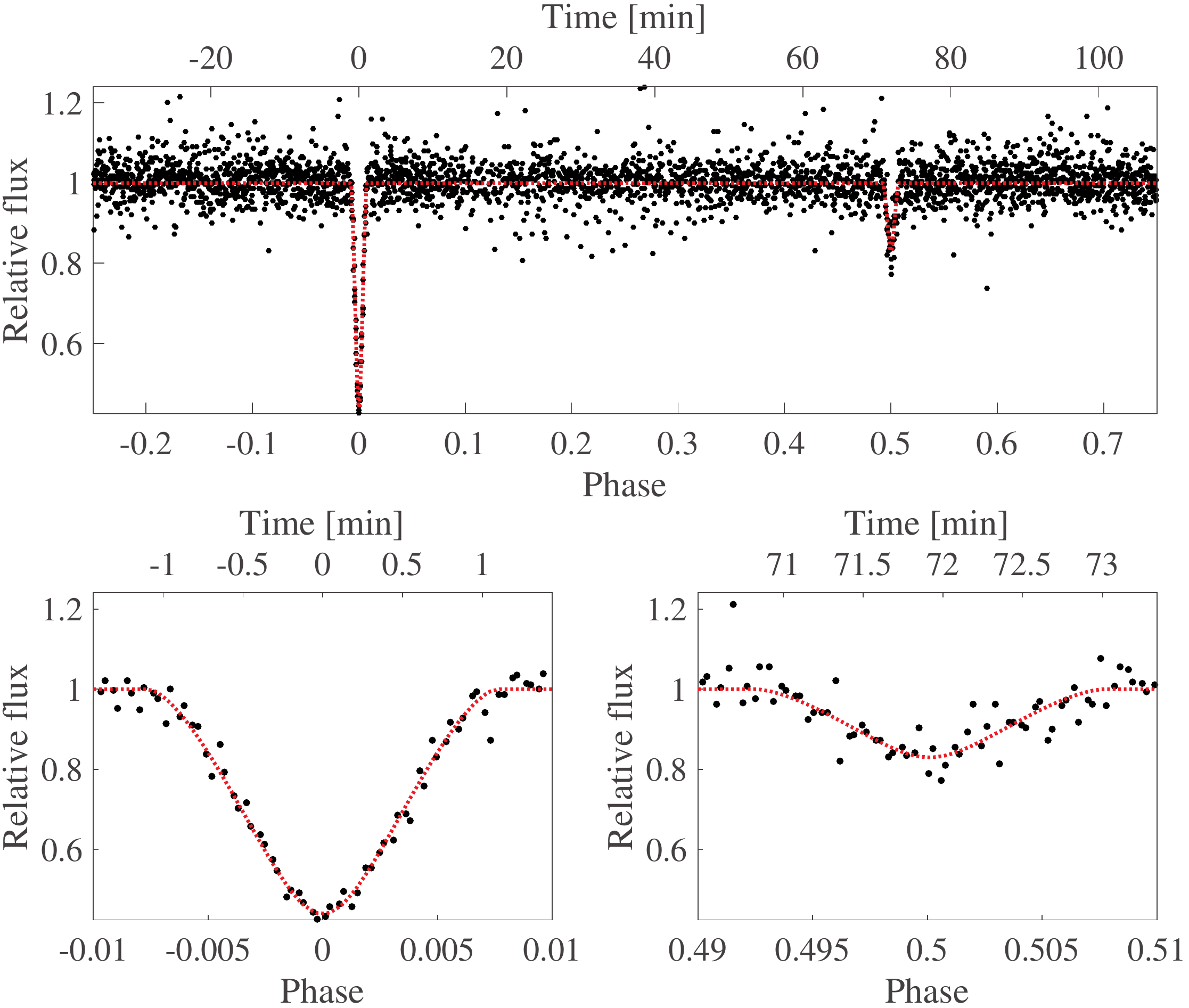}
\caption{Combined, phase-folded and normalized APO and McDonald light curve. The red dashed line represent the best-fitting model (see Table~\ref{tab:MCMCResults}). Top: full-period light curve; bottom left: zoom on the primary eclipse; bottom right: zoom on the secondary eclipse.}
\label{fig:LC}
\end{figure}

The two light curves were then synchronized by finding the time shift that minimizes the $\chi^2$ of the differences between the two data sets, with differences taken between APO data points and interpolated McDonald data points. The merged light curve was phase-folded according to the period found from the long-term \textit{K2} light curve. The zero phase (defined as the primary mid-eclipse) was determined using a parabolic fit around the middle of the primary eclipse. Fig.~\ref{fig:LC} shows the combined, phase-folded light curve from the fast photometry. We have verified that the light-curve areas within each eclipse region are consistent between the \textit{K2} and the ground-based light curves.

Next, we model the folded light curve during the two eclipses. Given the short orbital period of the system, tidal circularisation is likely.
We will therefore assume a circular orbit, but will test this assumption by allowing for a phase shift that differs by $\delta$ from 0.5 between the primary and secondary eclipses, due to either eccentricity of the orbit
or to a R{\o}mer delay \citep{Kaplan_2010}. The eclipse duration is related to the system parameters by
\begin{equation}
\Delta t_{\textrm{eclipse}} = P \frac{ \sqrt{\left( 1 + \frac{R_2}{R_1} \right) ^2 - b^2}}{\pi \frac{a}{R_1}},
\end{equation}
where $P$ is the orbital period, $R_1$ and $R_2$ are the primary and secondary radii, respectively, and $a$ is the semi-major axis. The impact parameter is defined as $b = \left(a/R_1\right) \cos i$, where $i$ is the orbital plane's inclination to the line of sight. Since the duration of the eclipse is very short compared to the period, a constant relative velocity between the two components during the eclipses can be assumed.

To account for limb-darkening we use the limb-darkening law of \citet{Claret+2000},
\begin{equation}\label{eq:LD}
\frac{I\left(\mu\right)}{I\left(1\right)} = 1 - c_1 \left(1 - \mu^\frac{1}{2}\right) - c_2 \left( 1 - \mu \right) - c_3 \left(1 - \mu^\frac{3}{2}\right) - c_4 \left(1 - \mu^2 \right),
\end{equation}
where $I$ is the specific intensity, $\mu$ is the cosine of the angle between the line of sight and the direction of the emergent flux from the center of the star, and $c_{1-4}$ are the limb-darkening coefficients. The limb-darkening coefficients for the Johnson-Kron-Cousins \textit{UBVRI}, and for Large Synoptic Survey Telescope (LSST) \textit{ugrizy} filters have been calculated by \citet{Gianninas+2013} for various WD effective temperatures and surface gravity values. From studying the BG40 transmission curve and the total instrument throughput of Agile and ProEM Camera, we find that a linear combination of the LSST \textit{ugr} filters,
\begin{equation}\label{eq:LD_filter}
\textrm{BG40Sys} = 0.11u+0.48g+0.41r,
\end{equation}
provides a reasonable approximation for the wavelength response of the BG40 filter combined with that of the detector and the instrument. For every pair of assumed values of $T_{\textrm{eff}}$ and $\log g$, the limb-darkening coefficients of each LSST filter were interpolated from the table of \citet{Gianninas+2013} using bilinear interpolation. In our analysis, each WD is modelled as a disc divided into $200$ concentric annuli. Every annulus is assigned a flux, corresponding to the limb-darkening formula on equation~(\ref{eq:LD}). The fluxes calculated for the different LSST filters are combined using equation~(\ref{eq:LD_filter}). A model eclipse light curve is then produced by means of numerical integration over the uneclipsed regions of each WD.

The light curve was modelled using a Matlab implementation\footnote{The code used here is a modified version of Aslak Grinsted's \texttt{gwmcmc}. The unmodified version can be found at \url{https://github.com/grinsted/gwmcmc}} of the Goodman and Weare Affine Invariant Markov chain Monte Carlo (MCMC) Ensemble sampler \citep{Goodman+2010,ForemanMackey+2013}. We used an MCMC run with 500 `walkers' and 100,000 steps. The MCMC routine minimizes the $\chi^2$ value of the fit, over a 5-dimensional parameter space, defined by the ratio of WD radii, $R_2/R_1$, the ratio of semi-major axis to primary radius, $a/R_1$, the impact parameter $b$, the ratio of the central surface brightness of the two WDs, $f_2/f_1$, and the secondary eclipse phase deviation, $\delta$. The orbital period is taken from the \textit{K2} light-curve analysis. To check for convergence of the MCMC process, we verified that the distribution of $\Delta \chi^2$ from the last $75\%$ of the steps is well approximated by a $\chi^2$ distribution. The medians and $\pm 1\sigma$ intervals of the parameter distributions in the last 75,500 steps of the MCMC routine (which are used to estimate the errors in the parameters), are listed in the second column of Table~\ref{tab:MCMCResults}. The third column lists the best-fitting model.

The limb-darkening coefficients (four for each WD) were kept fixed during each MCMC run. The coefficients (\textit{u}: 0.92, -0.93, 0.92, -0.34; \textit{g}: 0.90, -0.85, 0.72, -0.25; \textit{r}: 0.85, -0.96, 0.88, -0.32) were calculated using equation~(\ref{eq:LD_filter}), with the effective temperature and $\log g$ values found for the primary WD in a spectral analysis ($T_1 \sim 25000$\,K, $\log g_1 \sim 7.3$, see section~\ref{sec:Atmospheric} below), assigned to it. For the secondary, $T_2$ and $\log g_2$ values ($T_2 \sim 14000$\,K, $\log g_2 \sim 7.1$, derived in Section~\ref{sec:Combined} below) were used to evaluate the secondary WD limb-darkening coefficients (\textit{u}: 0.77, -0.74, 0.96, -0.38; \textit{g}: 0.66, -0.13, 0.28, -0.11; \textit{r}: 0.61, -0.14, 0.07, -0.03). Varying $T_{\textrm{eff}}$ and $\log g$ had little effect on the results.

\begin{figure*}
\includegraphics[width=0.85\textwidth]{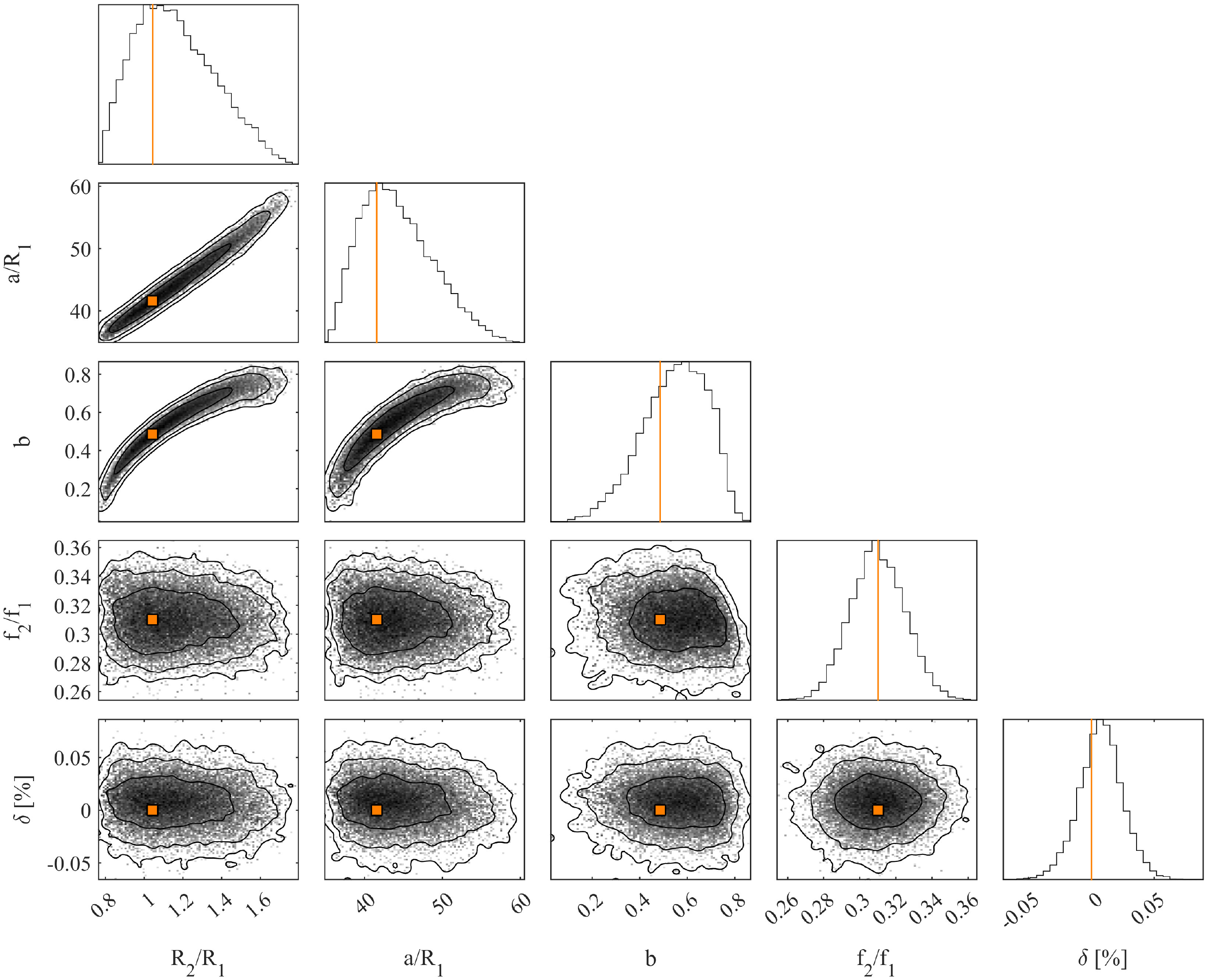}
\caption{One- and two-dimensional projections histograms of converged MCMC chains, from the modelling of the eclipse light curves. The contours indicate the 1, 2 and $3\sigma$ levels. Note the strong covariance between $R_2/R_1$, $a/R_1$ and $b$. The orange markings indicate the best-fitting model (see Table~\ref{tab:MCMCResults}).}
\label{fig:Cov}
\end{figure*}

Table~\ref{tab:MCMCResults} and Figs.~\ref{fig:LC} and \ref{fig:Cov} show the results of our modelling process. An acceptable best fit is found, with a reduced $\chi^2_R=1.4$. The phase deviation $\delta$ is consistent with zero. There is thus no indication of orbital eccentricity. From the full solution of the system parameters (see Section~\ref{sec:Combined}, below), the expected R{\o}mer delay is $\Delta t_\textrm{rom}=a/c \sin i\approx 2.5$\,s, or $\delta\approx 0.00029$, which is consistent, within the errors, with our result for $\delta$. The other parameters are determined to 10-20\% precision. The secondary WD has a radius comparable to that of the primary, but with a preference to be slightly larger. The secondary WD's lower central surface brightness means it is substantially cooler than the primary, as already clear from the ratio of eclipse depths. The ephemeris for mid-primary eclipse is $\textrm{BJD(TDB)}_\textrm{p}=2457127.801118\pm0.000005 + \left(0.099865\pm0.000013 \right) E$, where $E$ is the number of periods since UT 2015 April 15, 07:05:08.

\begin{table}
\centering
\caption{Eclipse data model results}
\label{tab:MCMCResults}
\def\arraystretch{1.5}
\begin{tabular}{l l l}
\hline
Parameter & MCMC results & \shortstack[l]{Best-fitting model \\ (see Fig.~\ref{fig:LC})} \\
\hline
$R_2/R_1$ & $1.14^{+0.24}_{-0.19}$  & $1.04$ \\
$a/R_1$ & $43.7^{+5.5}_{-4.1}$  & $41.6$ \\
$b$ & $0.56^{+0.13}_{-0.15}$ & $0.49$ \\
$f_2/f_1$ & $0.310\pm0.015$ & $0.310$ \\
$\delta$ & $0.00008\pm0.00017$ & $0.00010$ \\
$\chi^2_\textrm{min}$ & & $4414.7$ \\
Degrees of freedom & & $3049$ \\
\hline
\end{tabular}
\end{table}

We have further searched for the effects of relativistic beaming in the \textit{K2} light curve. The primary has a radial velocity amplitude of $K_1 \approx 230$\,km\,s$^{-1}$ (see Section~\ref{sec:Combined}), and its expected beaming relative semi-amplitude \citep{Loeb+2003,Zucker+2007} is $A\approx B K_1/c$, with a maximum at phase 0.25 (primary approaching us) and minimum at 0.75 (primary receding), where $B$ is a k-correction-like factor depending on the spectrum and on the observed bandpass. Integrating in wavelength over a Planck spectrum with the primary WD's temperature $T_1$ (see Section~\ref{sec:Atmospheric}) times the \textit{Kepler} response function, following \citet{Bloemen+2011}, we find $B=1.6$. For the secondary, with its lower temperature (see Section~\ref{sec:Combined}), we find  $B\approx 2.1$. Furthermore, the secondary has about 0.31 the surface brightness of the primary, but (see Section~\ref{sec:Combined}) a surface area larger by $(R_2/R_1)^2$, and a radial velocity larger by a factor $M_1/M_2$, but with an opposite sign. Accounting also for the uncertainties in all these parameters, the total expected beaming semi-amplitude is thus $A\approx 0.00055^{+0.00025}_{-0.00033}$, which could be quite small because of cancellation of the beaming effects of the primary and the secondary.

To search for a beaming signal, we have modelled the folded \textit{K2} light curve as a constant with two half-hour-long `boxcars' corresponding to the two eclipses, and modulated the model light curve with a sine function with the system's period and a maximum at phase 0.25. Varying the sine's amplitude, the best fit to the \textit{K2} data is with $A=0.00049\pm 0.00019$, a marginal $2.5\sigma$ detection of beaming in the \textit{K2} data, within the expected error range (see Fig.~\ref{fig:K2}). The eclipses around phases 0 and 0.5, when the beaming effect is zero, are brief, and therefore inclusion of the beaming effect does not influence the parameters derived above from the ground-based eclipse data analysis. We will ignore gravitational lensing effects, which are pronounced only during the eclipses, at levels of $<10^{-4}$, well below the precision of the ground-based data.

\subsection{Atmospheric model fit}\label{sec:Atmospheric}

\begin{figure}
\includegraphics[width=\columnwidth]{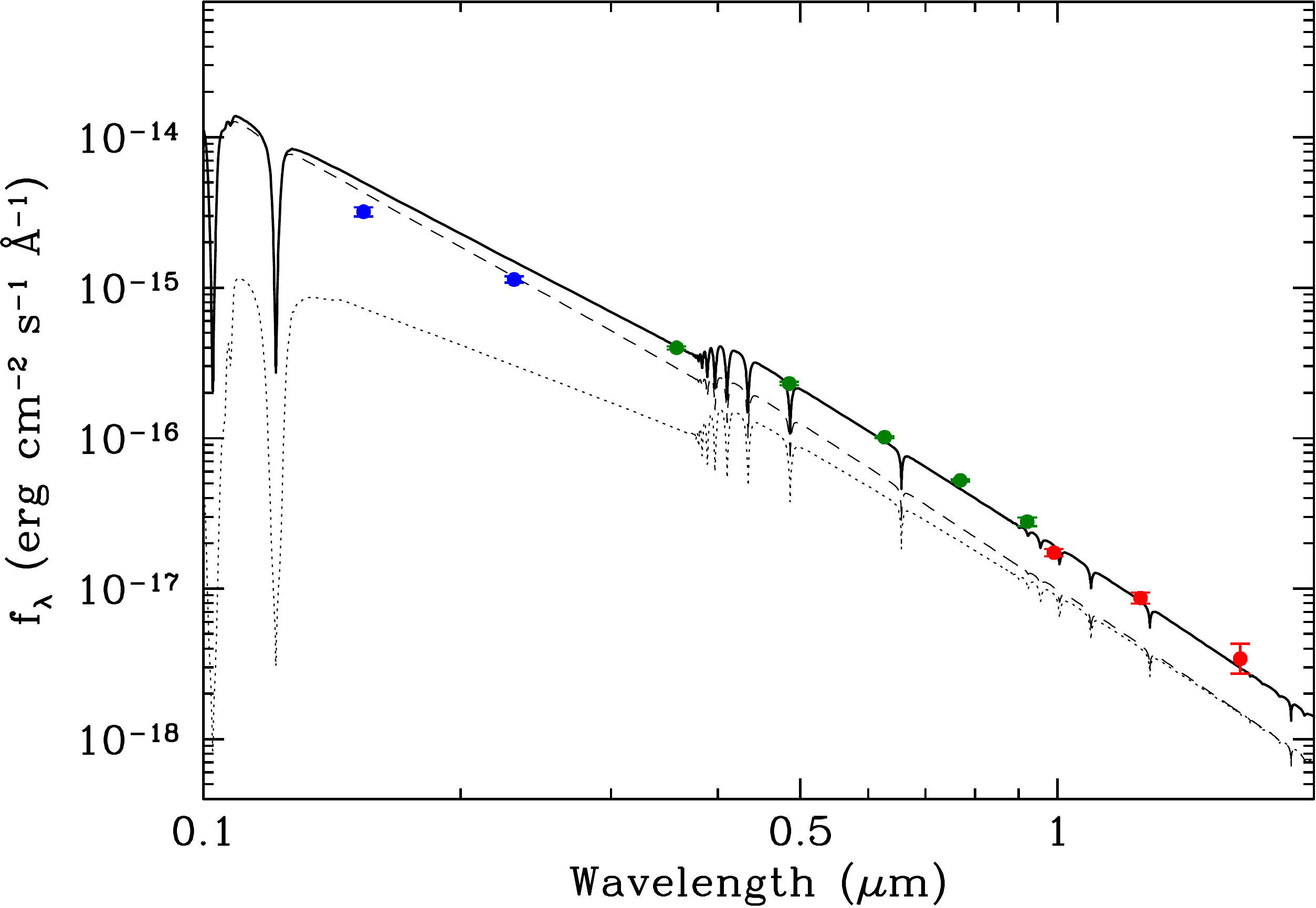}
\caption{Observed spectral energy distribution for SDSS~J1152+0248 compared to model spectra. Dots, from left to right, are \textit{GALEX} far-UV, \textit{GALEX} near-UV, \textit{SDSS} \textit{u}', \textit{g}', \textit{r}', \textit{i}' and \textit{z}' fluxes, and \textit{UKIDSS} Y, J and H fluxes. All fluxes have been corrected for Galactic extinction. Models are a $\sim 25,000$\,K primary DA WD (dashed line), a $\sim 14,000$\,K secondary DA WD (dotted line), and their sum (solid line).}
\label{fig:SED}
\end{figure}

\begin{figure}
\includegraphics[width=\columnwidth]{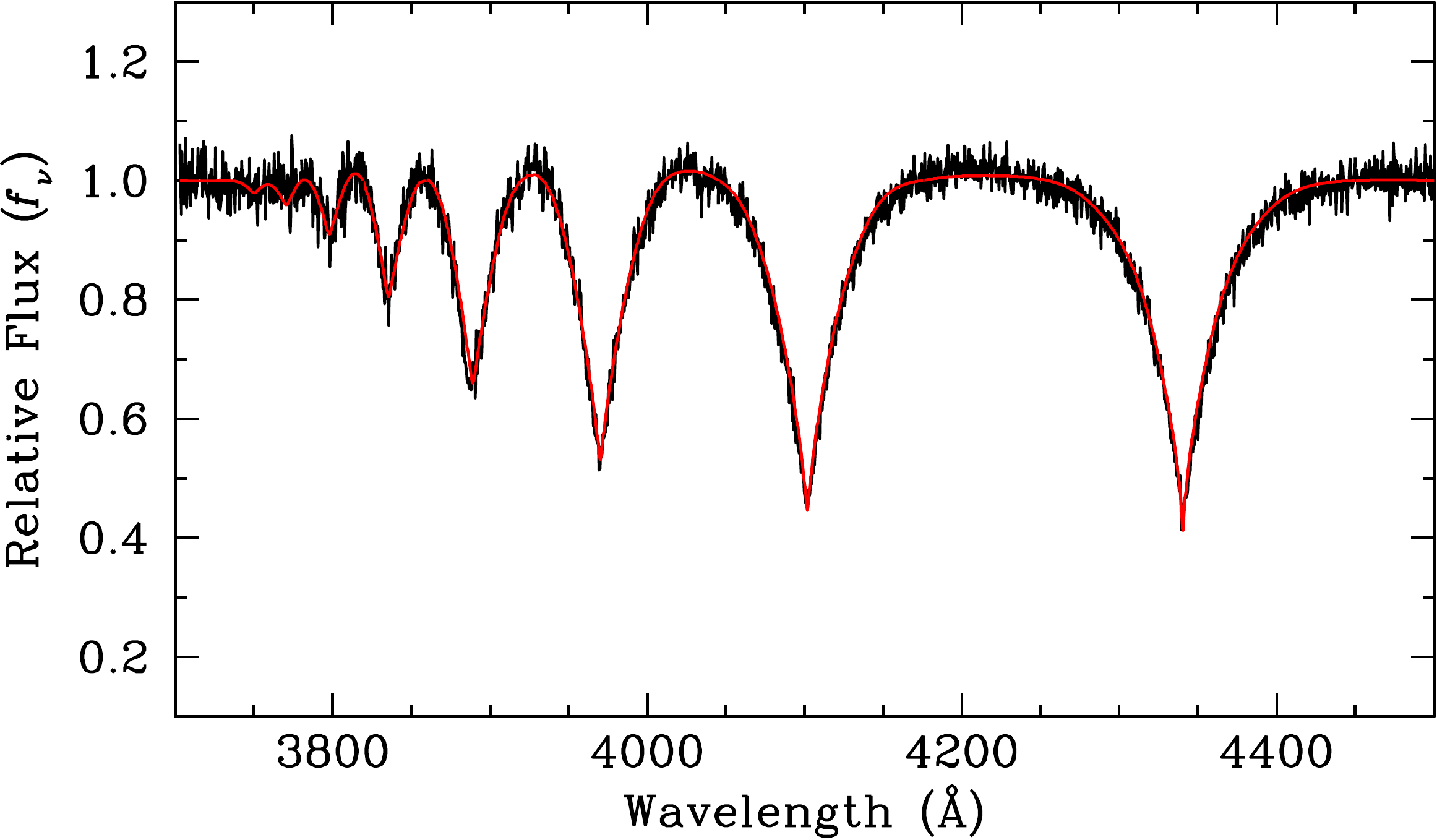}
\caption{Atmospheric model fit to the MMT spectrum (a sum of all exposures), consisting of a primary DA WD model with $T_1 = 25,500$\,K and $\log g_1 \sim 7.4$, contributing $\sim 0.8$ of the luminosity in this band, plus a $T_2 = 14,300$\,K, $\log g_2 \sim 7$ DA WD model that contributes $\sim 0.2$.}
\label{fig:Atmospheric}
\end{figure}

To further constrain the system, the spectra were fitted using WD atmospheric models \citep{Bergeron+1994,Tremblay+2009,Gianninas+2011}. By experimenting with the models, we concluded that there is no evidence for dilution of the Balmer lines of the primary by a featureless continuum from the cooler secondary, but rather that the secondary strengthens the lines in the total spectrum, indicating it is also a DA type. The ultraviolet-bright spectral energy distribution (SED) seen in this object's photometry from SDSS and from \textit{GALEX} (see Fig.~\ref{fig:SED}), indicates a hot primary WD. We therefore calculated the expected SED of a hot primary DA combined with a cooler DA model spectrum, scaled to have 0.18 to 0.32 of the total \textit{g}-band flux in the spectra. This range corresponds to $L_2/\left(L_1+L_2\right)$, the fractional contribution of the secondary to the luminosity from both WDs, considering the values of $f_2/f_1$ and $R_2/R_1$ from the eclipse light curve fit, as well as the different limb-darkening profiles in the $4000-5000\,\AA$ spectral region of the hotter and the cooler WDs. The primary effective temperature, $T_1$, was searched over a range of temperatures. For each value of $T_1$ the secondary effective temperature, $T_2$, was calculated using the measured central surface brightness ratio and the system response (see Sec~\ref{sec:Combined}). By comparing the models to the observed SED, we were able to constrain $T_1$ to $\sim 25,500$\,K, with $T_2 \sim 14,300$\,K. The somewhat poor fit of the GALEX FUV and NUV values (see Fig.~\ref{fig:SED}) might be explained by the uncertainty of the Galactic extinction values in these bands as estimated by \citealt{Yuan+2013}. Using different extinction values \citep{Seibert+2005,Wyder+2007} did not improve the results. Atmospheric model fits to the Balmer lines in the WHT and MMT spectra yield similar temperatures, as well as providing estimates of the surface gravity of the primary, $\log g_1$. The fitting results for the primary WD listed in Table~\ref{tab:Atmospheric}, are based on the sum of all 13 MMT spectra, after correcting all epochs to zero velocity (see Fig.~\ref{fig:Atmospheric}). The resulting values of $T_1$ and $\log g_1$ (as well as the secondary WD's $T_2$ and $\log g_2$ from our subsequent analysis, see Sec~\ref{sec:Combined}) were then used to recalculate the limb-darkening coefficients (see Sec~\ref{sec:Model}), and rerun the MCMC fit. The new results, which are those presented in Table~\ref{tab:MCMCResults}, were within $1\sigma$ of the results of the initial run.

\begin{table}
\centering
\caption{Primary WD model atmosphere fit parameters.}
\label{tab:Atmospheric}
\begin{tabular}{l l l}
\hline
Parameter & Value \\ \hline
Primary spectral type & DA \\
$T_1$ (K) &  $25500\pm1000$ \\
$\log g_1$ & $7.42\pm0.25$ \\
\hline
\end{tabular}
\end{table}

\subsection{Radial-velocity curves}\label{sec:RV}

\begin{figure}
\includegraphics[width=\columnwidth]{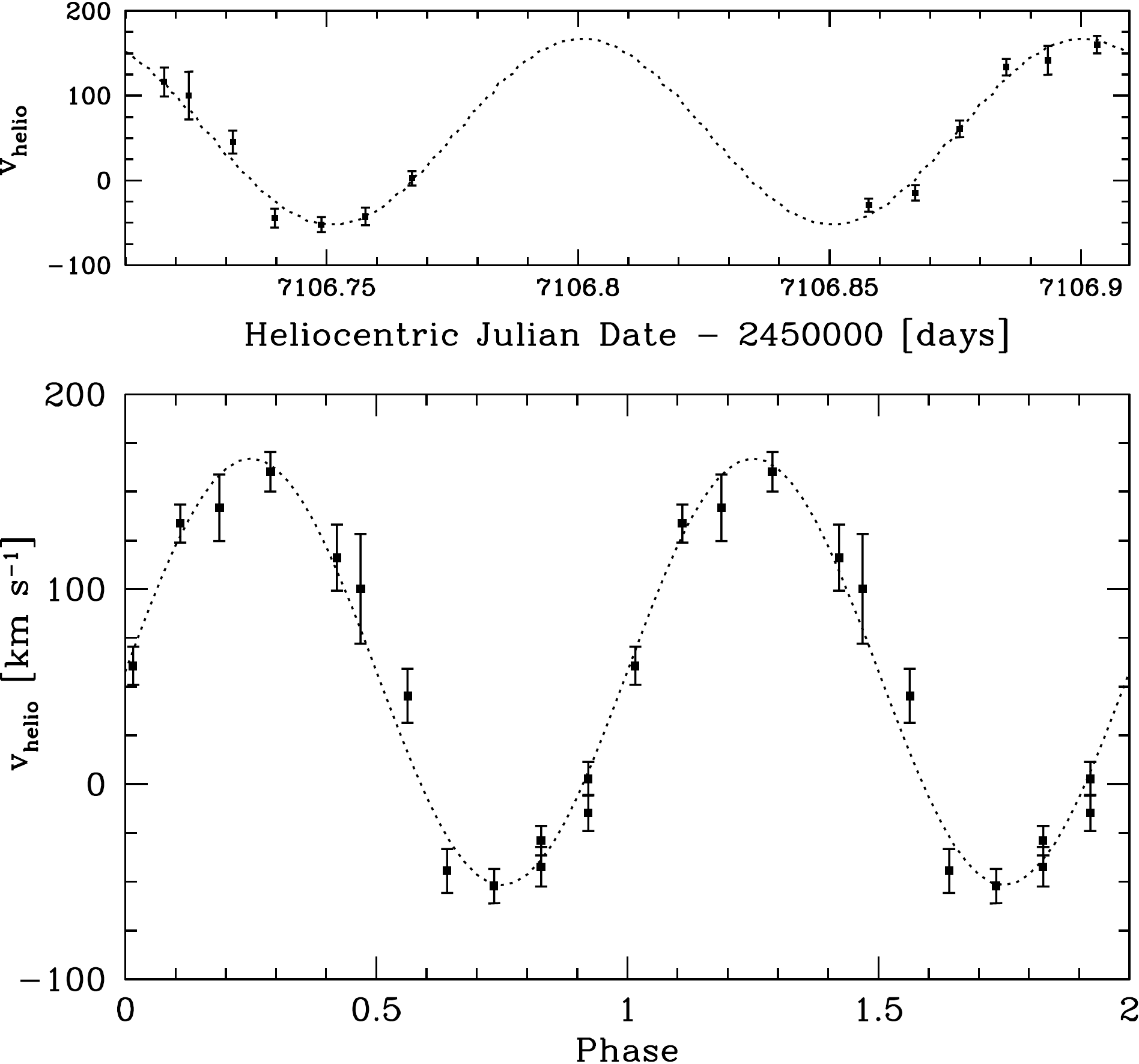}
\caption{Radial velocity data and model fit. Top panel: radial velocity vs. time; bottom panel: phase-folded radial velocity curve. The dashed line is the best-fitting model.}
\label{fig:RV}
\end{figure}

The kinematic data for SDSS~J1152+0248 were modelled as follows. We measured radial velocities using the cross-correlation package \texttt{RVSAO} \citep{Kurtz+1999}. We used H$\gamma$ and higher order Balmer lines from the MMT spectra for our velocity measurements, with a median error per epoch of $10$\,km\,s$^{-1}$. The measured radial velocity (RV) curve was then fit with a sine function, assuming the measured \textit{K2} photometric period, and with the measured amplitude $K_\textrm{meas}$, and the mean velocity $\gamma$ as free parameters:
\begin{equation}
v_r\left(\varphi\right) = \gamma + K_\textrm{meas} \sin \left(2\pi\varphi\right),
\end{equation}
where $\varphi$ is the orbital phase. The best-fitting model gives $K_\textrm{meas} = 109 \pm$ 5\,km\,s$^{-1}$ and $\gamma = 55\pm 5$\,km\,s$^{-1}$ (see Fig.~\ref{fig:RV}).

\subsection{Combined analysis of system parameters}\label{sec:Combined}

We now combine the results from our previous analyses: the period $P$ from the \textit{K2} photometry; the ratios $R_2/R_1$, $a/R_1$, $b$, $f_2/f_1$ and $\delta$ from the fast-photometry eclipse light-curve analysis; the primary effective temperature $T_1$ and surface gravity $\log g_1$ from the model atmosphere fit, and the measured RV amplitude $K_\textrm{meas}$ from the cross-correlation analysis, to estimate the properties of both WDs and the orbital separation (see Table~\ref{tab:Results}).

In order to derive the secondary WD temperature, $T_2$, we computed the ratio between two Planck functions, each integrated over the filter plus system response (see section \ref{sec:Model}), for various temperatures and $T_1$, to find the temperature that gives a ratio corresponding to $f_2/f_1=0.310\pm0.015$. This results in a secondary WD temperature $T_2=14350^{+500}_{-490}$\,K.

\begin{figure}
\includegraphics[width=\columnwidth]{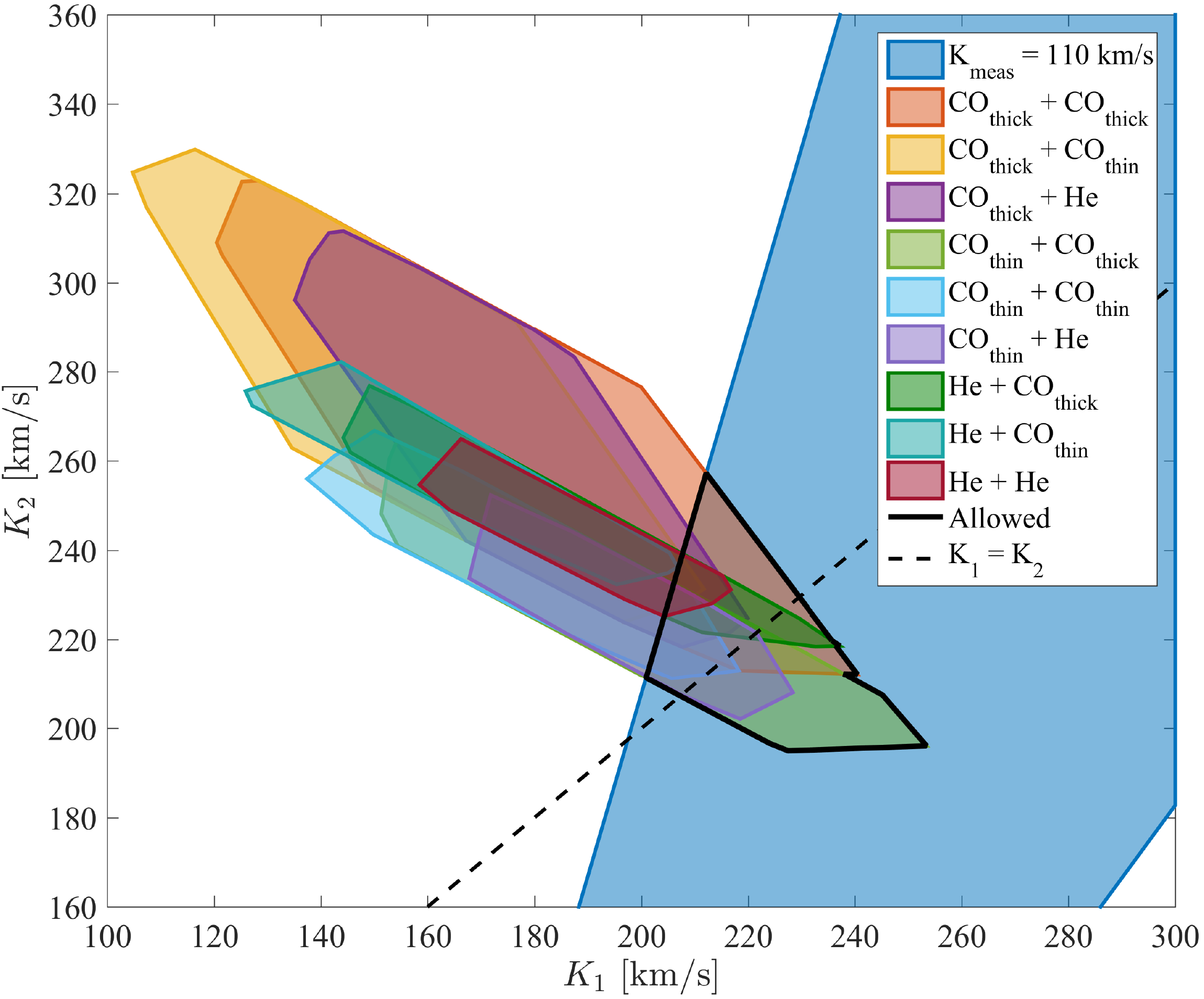}
\caption{Constraints on the RV amplitude ($K_1$, $K_2$) plane. The ``allowed'' area for all possible core compositions is represented by the black contour. The $K_\textrm{meas}=110$\,km\,s$^{-1}$ constraint polygon is derived from the cross-correlation simulation (see Section~\ref{sec:Combined}), while the other polygons represent the $R_2/R_1$ and $a/R_1$ constraints for each core composition combination.}
\label{fig:K}
\end{figure}

Next, we tested for the possibility that $K_\textrm{meas}$ is being biased low  in the cross-correlation analysis by the effect of the (undetected) lines from the secondary WD. To do this, we simulated a model spectrum consisting of a hot DA WD combined with a cooler DA WD, with luminosity ratios in the range constrained by the light curve analysis, and with a range of RV phases for each of the two components. Each model spectrum was noised at the $5\%$ level and then cross-correlated with a model single hot DA WD as in the real observations, for the same orbital phase values as the MMT observations, to simulate a ``measured'' RV amplitude. We repeated the calculation over a grid of primary and secondary RV amplitudes ($K_1$ and $K_2$, respectively), for the minimal and maximal values of the fractional contribution of the secondary to the luminosity from both WDs. The required constraint of a simulated ``measured'' RV amplitude of $K_\textrm{meas}=110\pm5$\,km\,s$^{-1}$ is shown as the broad blue swath in Figure~\ref{fig:K}.

To derive the possible mass combinations, allowed by the constraints implied by the light curve and the RV curve, we have taken a grid of $M_{1,2}$ masses, between $0.2$ and $1.4\,M_{\sun}$, and calculated the corresponding $R_{1,2}$ radii using theoretical WD evolutionary cooling sequences. For a WD of given mass and given core and atmosphere compositions at birth, these models calculate the radius as a function of the declining temperature. Lacking prior  knowledge on the WD core composition and hydrogen envelope thickness, we have considered a range of models: carbon-oxygen cores with `thin' ($10^{-10}$ mass fraction) or `thick' ($10^{-4}$ mass fraction) hydrogen envelopes and with a $10^{-2}$ mass fraction helium envelope \citep{Fontaine+2001}\footnote{\url{http://www.astro.umontreal.ca/~bergeron/CoolingModels/}}, and helium core WDs \citep{Althaus+2013}\footnote{\url{http://cdsarc.u-strasbg.fr/viz-bin/qcat?J/A+A/557/A19}}. We only considered helium core models with $R < 0.08\,R_{\sun}$, since the densities indicated by the period and the eclipse durations imply that both WDs are degenerate. For each combination of core compositions of the primary and the secondary, $R_{1,2}$ were estimated by interpolating over temperature and mass in the cooling-sequence tables, using the derived $T_1$ and $T_2$ temperatures. ($M_1$, $M_2$) points which gave a radius ratio, $R_2/R_1$, within the limits derived from the light curve (see Table~\ref{tab:MCMCResults}), were used to delineate the allowed region in the ($M_1$, $M_2$) plane.

Similarly, to further constrain ($M_1$, $M_2$), we used Kepler's law to calculate the expected orbital separation over the ($M_1$, $M_2$) grid,
\begin{equation}
a = \left( \frac{P^2}{4 \pi^2} G \left( M_1 + M_2 \right) \right)^{\frac{1}{3}}
\end{equation}
where $P$ is the orbital period measured from the \textit{K2} light curve. Again, only grid points with an $a/R_1$ value within the limits derived from the light curve (see Table~\ref{tab:MCMCResults}), were included in the allowed region of the ($M_1$, $M_2$) plane.

\begin{figure}
\includegraphics[width=\columnwidth]{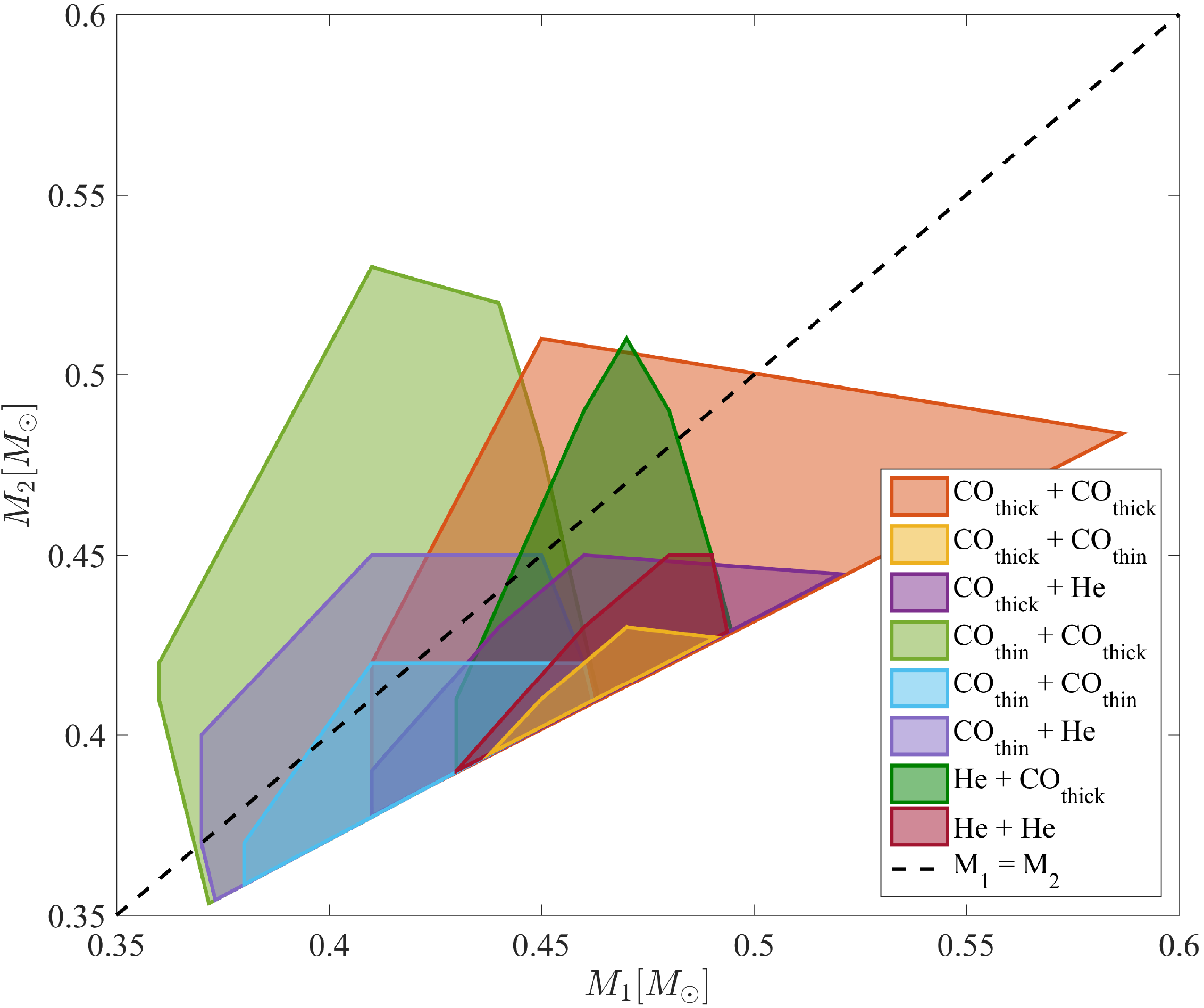}
\caption{Constraints on the mass ($M_1$, $M_2$) plane. Each polygon represent a different primary+secondary core composition combination. Note that the $\textrm{He}+\textrm{CO}_\textrm{thin}$ option is ruled out by the constraints.}
\label{fig:M}
\end{figure}

We then calculated the expected RV amplitudes, $K_{1,2}$, over the same ($M_1$, $M_2$) grid, using
\begin{equation}
K_{1,2} = \sqrt{\frac{GM_{2,1}}{a\left(1 - e^2\right)}}\sqrt{\frac{M_{2,1}}{M_1+M_2}}\sin i
\end{equation}
where the eccentricity, $e$, was taken as zero, and the inclination, $i$, was taken from the MCMC results (see Table~\ref{tab:MCMCResults}), to translate the $R_2/R_1$ and $a/R_1$ constraints on the ($M_1$, $M_2$) plane to constraints on the ($K_1$, $K_2$) plane. Combining these constraints with the constraint on $K_\textrm{meas}$, derived from the cross-correlation simulation, as shown in Figure~\ref{fig:K}, we see that, despite the measured cross-correlation amplitude of $\sim 110$\,km\,s$^{-1}$, the true RV amplitudes $K_{1,2}$ are actually $\sim 200-250$\,km\,s$^{-1}$. This can be considered an indirect detection of the effect on the lines from the secondary on the total observed spectrum. Finally, we transform the allowed region in the ($K_1$, $K_2$) plane back to the to the ($M_1$, $M_2$) plane, using
\begin{equation}
M_{1,2} = \frac{P \left( K_1 + K_2 \right)^2 \sqrt{1-e^2}}{2 \pi G \sin^3 i} K_{2,1},
\end{equation}
with the results shown in Figure~\ref{fig:M}.

Taking into account all possible core compositions, we obtain $M_1=0.47\pm0.11\,M_{\sun}$ for the mass of the primary WD, and $M_2=0.44\pm0.09\,M_{\sun}$ for the secondary WD (see Table~\ref{tab:Evolutionary} and Figures~\ref{fig:K} and \ref{fig:M}). The combination of a He-core primary WD and a `thick' CO-core secondary WD, is the only option which is ruled out by the combined constraints.

The effective temperature and mass were then used to estimate each WD's radius, $R_{1,2}$, and cooling age, $t_{1,2}$, using the theoretical WD evolutionary cooling sequences. The errors on $M_{1,2}$, $R_{1,2}$ and $t_{1,2}$ listed in Table~\ref{tab:Results} include both the statistical errors that propagate from $T_{1,2}$, and the theoretical uncertainty from the range of cooling models included. As seen in Table~\ref{tab:Evolutionary}, the model atmosphere fitting, combined with the WD evolutionary cooling sequences and the constraints derived from the light curve and the RV measurements, indicate both WDs are of low-mass, with $M_1 \approx 0.47\,M_{\sun}$ and $M_2 \approx 0.44\,M_{\sun}$. Both WDs have radii of $\sim 0.02\,R_{\sun}$. The cooling age of the primary is $\approx 52$\,Myr, while that of the secondary is $\approx 230$\,Myr.

\begin{table*}
\centering
\caption{Primary and secondary mass, radius and cooling age from WD evolutionary cooling sequences, constrained by the RV and light curves. References: CO core models \citep{Fontaine+2001}, He core models \citep{Althaus+2013}.}
\label{tab:Evolutionary}
\def\arraystretch{1.5}
\begin{tabular}{l l l l l l l l}
\hline
Primary type &
Secondary type &
$M_1$ ($M_{\sun}$) &
$M_2$ ($M_{\sun}$) &
$R_1$ ($R_{\sun}$) &
$R_2$ ($R_{\sun}$) &
$t_1$ (Myr) &
$t_2$ (Myr) \\
\hline
He & He &
$0.462\pm0.032$ &
$0.420\pm0.030$ &
$0.0193^{+0.0023}_{-0.0017}$ &
$0.0186^{+0.0015}_{-0.0013}$ &
$61^{+28}_{-23}$ &
$304^{+26}_{-51}$ \\

He & CO, thin &
--- &
--- &
--- &
--- &
--- &
--- \\

He & CO, thick &
$0.462\pm0.032$ &
$0.450\pm0.060$ &
$0.0193^{+0.0023}_{-0.0017}$ &
$0.0217^{+0.0038}_{-0.0042}$ &
$61^{+28}_{-24}$ &
$191^{+85}_{-52}$ \\

CO, thin & He &
$0.416\pm0.046$ &
$0.402\pm0.048$ &
$0.0194^{+0.0018}_{-0.0019}$ &
$0.0196^{+0.0024}_{-0.0021}$ &
$20.3^{+10.3}_{-4.0}$ &
$279^{+45}_{-61}$ \\

CO, thin & CO, thin &
$0.421\pm0.041$ &
$0.389\pm0.031$ &
$0.0192^{+0.0017}_{-0.0018}$ &
$0.0189^{+0.0029}_{-0.0021}$ &
$20.6^{+10.7}_{-4.0}$ &
$151^{+62}_{-24}$ \\

CO, thin & CO, thick &
$0.412\pm0.052$ &
$0.442\pm0.088$ &
$0.0195^{+0.0019}_{-0.0020}$ &
$0.0220^{+0.0044}_{-0.0047}$ &
$20.0^{+10.3}_{-4.0}$ &
$185^{+100}_{-63}$ \\

CO, thick & He &
$0.465\pm0.055$ &
$0.414\pm0.036$ &
$0.0205^{+0.0027}_{-0.0027}$ &
$0.0189^{+0.0018}_{-0.0015}$ &
$23.5^{+20.2}_{-6.9}$ &
$297^{+31}_{-57}$ \\

CO, thick & CO, thin &
$0.464\pm0.026$ &
$0.412\pm0.018$ &
$0.0203^{+0.0024}_{-0.0020}$ &
$0.0179^{+0.0032}_{-0.0017}$ &
$23.6^{+18.9}_{-6.8}$ &
$162^{+72}_{-24}$ \\

CO, thick & CO, thick &
$0.498\pm0.088$ &
$0.444\pm0.066$ &
$0.0195^{+0.0034}_{-0.0033}$ &
$0.0219^{+0.0039}_{-0.0043}$ &
$28^{+27}_{-11}$ &
$186^{+86}_{-54}$ \\

\hline
\multicolumn{2}{l}{Full range} &
$0.47\pm0.11$ &
$0.442\pm0.088$ &
$0.0197\pm0.0035$ &
$0.0213\pm0.0051$ &
$52\pm36$ &
$230\pm100$ \\

\hline
\end{tabular}
\end{table*}

The secondary WD radius was recalculated using the radii ratio derived from the MCMC, and $R_1$. The surface gravity of each WD was then calculated using $g_{1,2} = GM_{1,2}/R_{1,2}^2$. All the derived parameters ($R_2$, $a$, $i$, $T_2$, $\log g_1$, $\log g_2$) were calculated using the MCMC chains, while normally distributed chains of the same length were created for non-MCMC parameters. The medians and $1\sigma$ intervals of the derived parameters are listed in Table~\ref{tab:Results}.

The time till merger of the system, due to gravitational wave losses, calculated using
\begin{equation}
t_\textrm{merger} = \frac{5}{256}\frac{c^5}{G^3}\frac{a^4}{M_1 M_2 \left(M_1 + M_2 \right)},
\end{equation}
is $460^{+720}_{-280}$\,Myr.

A distance to the system, $464^{+90}_{-85}$\,pc, was estimated using the \textit{SDSS} \textit{g}-filter magnitude ($g=18.35$\,mag) and a $0.08$\,mag Galactic extinction correction, compared to the summed \textit{g}-band luminosities obtained from the derived temperatures and radii.

\begin{table}
\centering
\caption{System parameters}
\label{tab:Results}
\def\arraystretch{1.5}
\begin{tabular}{l l l}
\hline
Parameter & Value & Source \\
\hline
$P$ (h) & $2.39677\pm0.00031$ & \textit{K2} light curve\\
$K_\textrm{meas}$ (km\,s$^{-1}$) & $109\pm5$ & RV\\
$K_1$ (km\,s$^{-1}$) & $227\pm26$ & Constrained simulated RV\\
$K_2$ (km\,s$^{-1}$) & $226\pm31$ & Constrained simulated RV\\
$M_1$ ($M_{\sun}$) & $0.47\pm0.11$ & Constrained models \\
$M_2$ ($M_{\sun}$) & $0.442\pm0.088$ & Constrained models \\
$R_1$ ($R_{\sun}$) & $0.0197\pm0.0035$ & Evolutionary sequences \\
$R_2$ ($R_{\sun}$) & $0.0223^{+0.0064}_{-0.0050}$ & Derived \\
$a$ ($R_{\sun}$) & $0.86^{+0.19}_{-0.17}$ & Derived \\
$e$ & $\sim 0$ & Derived \\
$i$ (\degr) & $89.275^{+0.140}_{-0.081}$ & Derived \\
$T_1$ (K) &  $25500\pm1000$ & SED\\
$T_2$ (K) & $14350^{+500}_{-490}$ & Derived \\
$\log g_1$ & $7.52\pm0.19$ & Derived \\
$\log g_2$ & $7.38\pm0.24$ & Derived \\
$t_1$ (Myr) & $52\pm36$ & Evolutionary sequences \\
$t_2$ (Myr) & $230\pm100$ & Evolutionary sequences \\
$t_\textrm{merger}$ (Myr) & $460^{+720}_{-280}$ &  Derived \\
$D$ (pc) & $464^{+90}_{-85}$ &  Derived \\
\hline
$T_0$ (BJD(TDB)) & \multicolumn{2}{l}{$\begin{aligned}
2457127.801118\pm0.000005 \\
+ \left(0.099865\pm0.000013 \right) E
\end{aligned}$} \\
\hline
\end{tabular}
\end{table}

\section{Discussion}

\begin{table*}
\centering
\caption{Known eclipsing double WD parameters. Sources: NLTT 11748 \citep{Kaplan+2014}, CSS 41177 \citep{Bours+2014,Bours+2015}, GALEX~J1717+6757 \citep[GALEX~J171708.5+675712,][]{Vennes+2011,Hermes+2014}, SDSS~J0651+2844 \citep[SDSS~J065133.338+284423.37,][]{Hermes+2012}, SDSS~J0751-0141 \citep[SDSS~J075141.18-014120.9,][]{Kilic+2014}, SDSS~J1152+0248 this work.  *Estimated from non-detection of a secondary eclipse in \citep{Kilic+2014}.}
\label{tab:EB}
\def\arraystretch{1.5}
\begin{tabular}{l l l l l l l}
\hline
Parameter & NLTT~11748 & CSS~41177 & GALEX~J1717+6757 & SDSS~J0651+2844  & SDSS~J0751-0141 & SDSS~J1152+0248 \\
\hline
Type & DA+D? & DA+DA & DAZ+D? & DA+D? & DA+D? & DA+D? \\

$M_1$ ($M_{\sun}$) &
$0.15\pm0.02$ &
$0.378\pm0.023$ &
$0.185\pm0.010$ &
$0.26\pm0.04$ &
$0.194\pm0.006$ &
$0.47\pm0.11$\\

$M_2$ ($M_{\sun}$) &
$0.72\pm0.02$ &
$0.316\pm0.011$ &
$\gtrsim 0.86$ &
$0.50\pm0.04$ &
$0.97^{+0.06}_{-0.01}$ &
$0.442\pm0.088$\\

$R_1$ ($R_{\sun}$) &
$0.0428\pm0.0009$ &
$0.02224\pm0.00041$ &
$0.093\pm0.013$ &
$0.0371\pm0.0012$ &
$0.155\pm0.020$ &
$0.0197\pm0.0035$\\

$R_2$ ($R_{\sun}$) &
$0.0109\pm0.0002$ &
$0.02066\pm0.00042$ &
$0.0092\pm0.0026$ &
$0.0142\pm0.0010$ &
$0.0092\pm0.0026$ &
$0.0223^{+0.0064}_{-0.0050}$\\

$\log g_1$ &
$6.35\pm0.06$ &
$7.322\pm0.015$ &
$5.67\pm0.05$ &
$6.76\pm0.04$ &
$5.54\pm0.05$ &
$7.52\pm0.19$\\

$\log g_2$ &
$8.22\pm0.01$ &
$7.305\pm0.011$ &
&
&
&
$7.38\pm0.24$\\

$T_1$ (K) &
$8706\pm137$ &
$22439\pm59$ &
$14900\pm200$ &
$16530\pm200$ &
$15750\pm240$ &
$25500\pm1000$ \\

$T_2$ (K) &
$7594\pm123$ &
$10876\pm32$ &
$\sim 16750\pm3050$ &
$8700\pm500$ &
$< 8900$* &
$14350^{+500}_{-490}$ \\

$P$ (h) &
$5.64145164(7)$ &
$2.784370445(36)$ &
$5.907288(72)$ &
$0.212557373(15)$ &
$1.85(99)$ &
$2.39677(31)$ \\

$a$ ($R_{\sun}$) &
$1.53\pm0.03$ &
$0.886\pm0.014$ &
&
&
&
$0.86^{+0.19}_{-0.17}$\\

$i$ (\degr) &
$89.6\pm0.1$ &
$88.97\pm0.02$ &
$\sim 86.8\pm0.6$ &
$84.4\pm2.3$ &
$85.4^{+4.2}_{-9.4}$ &
$89.275^{+0.140}_{-0.081}$\\

\hline
\end{tabular}
\end{table*}

\begin{figure}
\includegraphics[width=\columnwidth]{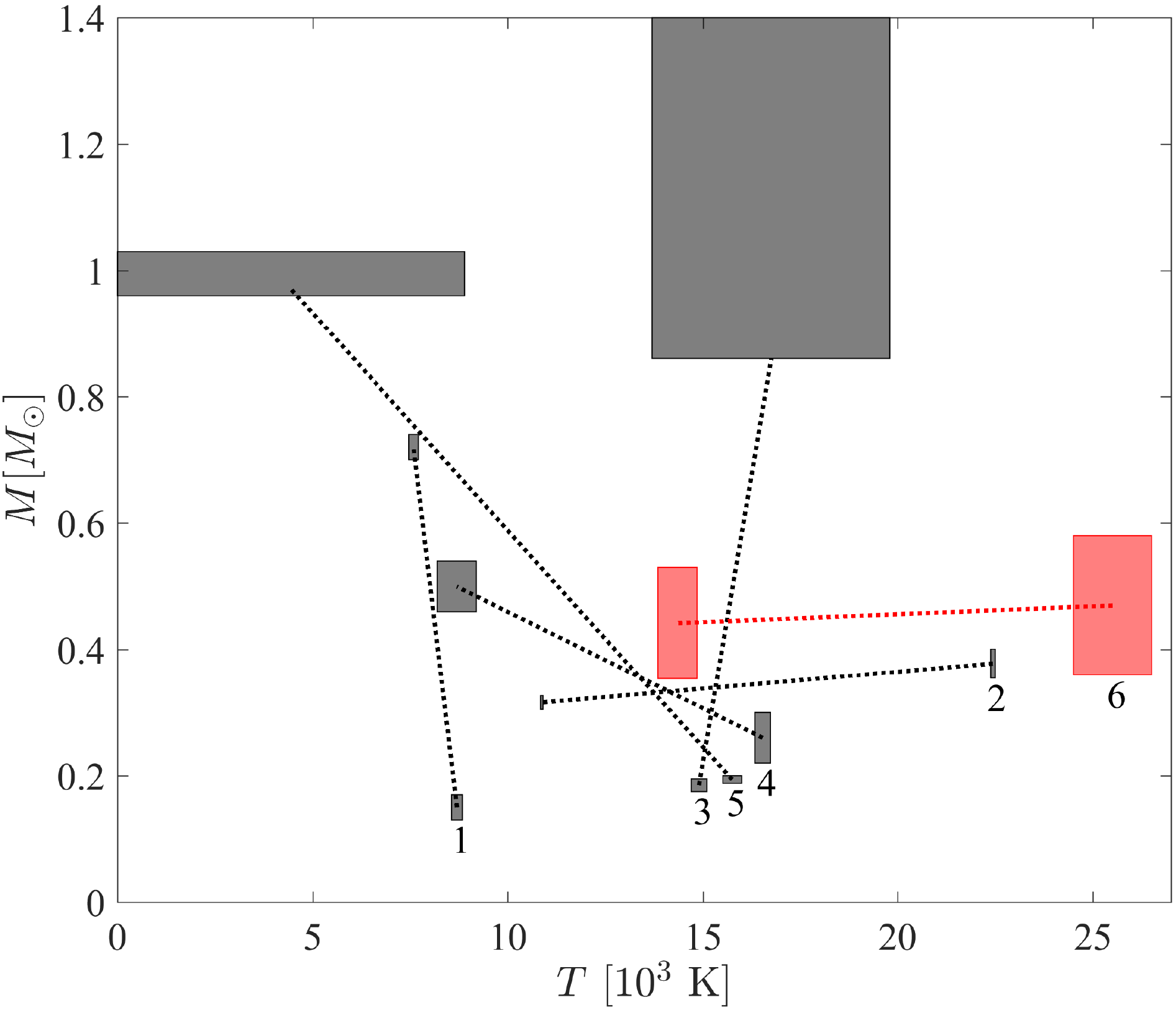}
\caption{Masses and temperatures of the components of known eclipsing double WDs. The labels mark the primary WD of each system: 1.~NLTT~11748, 2.~CSS~41177, 3.~GALEX~J1717+6757, 4.~SDSS~J0651+2844, 5.~SDSS~J0751-0141, 6.~SDSS~J1152+0248 (see Table~\ref{tab:EB} for references).}
\label{fig:EB}
\end{figure}

Comparing SDSS~J1152+0248 to the five previously known eclipsing double WDs (see Table~\ref{tab:EB} and Fig.~\ref{fig:EB}), we see that all of the systems consist of a primary DA \citep[specifically a DAZ in the case of GALEX~J171708.5+675712, see][]{Vennes+2011,Hermes+2014} and a secondary WD in a close orbit (with periods ranging from $\sim 12$\,min to $\sim 6$\,h). In three of the systems, the primary is an Extremely Low Mass (ELM) WD (with a mass $< 0.25\,M_{\sun}$), accompanied by a massive WD ($> 0.7\,M_{\sun}$). In the three remaining systems both WD masses are low and comparable, as is the case in SDSS~J1152+0248. CSS~41177, the only double-lined system among the six, is almost a twin to SDSS~J1152+0248. Both have a period of $\sim 2$\,h, a primary DA of mass $\sim 0.4\,M_{\sun}$ and temperature $\sim 2 \times 10^4$\,K, and a somewhat lighter secondary with a temperature of $\sim 1 \times 10^4$\,K.

\citet{Badenes+2012} have shown that $\sim 5\%$ of WDs are in short-period `double degenerate' binaries ($a \leq 0.05$\,AU $\sim 11\,R_{\sun}$). Assuming a separation distribution with equal numbers per logarithmic interval \citep{Badenes+2012}, the fraction of WDs with $a\lesssim 1 R_{\sun}$, like SDSS~J1152+0248, is $\sim 2.5\%$. Since the probability to observe a system such as SDSS~J1152+0248 at an inclination such that it undergoes eclipses is
\begin{equation}
p_{\textrm{eclipse}} = \frac{R_1 + R_2}{a} \approx 5\%,
\end{equation}
the probability, per WD, of discovering an eclipsing system is $\sim0.15\%$. As mentioned in Section~\ref{sec:Intro}, SDSS~J1152+0248 was discovered while searching for period variations in the light curves of $\sim 150$ WDs in \textit{K2} data, among which $\sim 125$ had signal-to-noise ratio sufficient for detecting the eclipse amplitudes seen in this system. Thus, we may have been mildly fortunate in discovering one such system in the \textit{K2} data, but the very-short-period double-WD fraction could perhaps be somewhat higher than the few percent estimated above.

\section{Conclusions}

We have discovered that SDSS~J1152+0248 is an eclipsing double WD, and have used spectroscopy and fast photometry to determine the parameters of the system. This short-period system consists of a low-mass hot primary DA WD, in a circular orbit with a cooler and somewhat lower-mass DA or DC WD with undetected lines, but likely also a DA. As in four of the five previously known systems of this type, the non-detection of spectral features of the secondary WD prevents a model-independent estimation of the physical parameters. Nevertheless, these systems permit measuring the masses, radii and temperatures of the secondary WD components, sharpening our picture of the close double WD binary population. Continued \textit{K2} observations of WDs in additional fields will likely uncover additional eclipsing double WDs, and will thus increase the known numbers of these useful systems.

\section*{Acknowledgements}
We thank Shai Kaspi for his help with the observations at Wise. We are grateful for valuable comments by J.J. Hermes and by the anonymous referee.
This work was supported in part by Grant 1829/12 of the I-CORE program of the PBC and the Israel Science Foundation (D.M. and T.M.). D.M. and T.M. acknowledge further support by individual grants from the ISF. The research by T.M. leading to these results has received funding from the European Research Council under the EU's Seventh Framework Programme (FP7/(2007-2013)/ERC Grant Agreement No.~291352). M.K., A.G. and K.B. gratefully acknowledge the support of the NSF under grant AST-1312678. M.K. and A.G. also acknowledge the support of NASA under grant NNX14AF65G. A.L. acknowledges support by the Raymond and Beverly Sackler Tel-Aviv University -- Harvard/ITC Astronomy Program. The William Herschel Telescope is operated on the island of La Palma by the Isaac Newton Group in the Spanish Observatorio del Roque de los Muchachos of the Instituto de Astrof\'{i}sica de Canarias. The analysis presented in this paper is based on observations obtained with the Apache Point Observatory 3.5-meter telescope, which is owned and operated by the Astrophysical Research Consortium. This paper includes data taken at the McDonald Observatory of The University of Texas at Austin. The authors acknowledge the Texas Advanced Computing Center (TACC)\footnote{\url{http://www.tacc.utexas.edu}} at The University of Texas at Austin for providing database resources that have contributed to the research results reported within this paper.




\bibliographystyle{mnras}
\bibliography{SDSSJ1152} 



%
%
%


\bsp	
\label{lastpage}
\end{document}